\documentclass{article}
\usepackage{mlspconf,amsmath,amsfonts,amssymb,amsthm,graphicx}
\usepackage{url}
\usepackage{hyperref}
\usepackage{caption}
\usepackage{scalefnt}
\usepackage{multirow}

\copyrightnotice{\begin{minipage}{\textwidth}
\scriptsize\copyright 2024 IEEE. Personal use of this material is permitted. Permission from IEEE must be obtained for all other uses, in any current or future media, including reprinting/republishing this material for advertising or promotional purposes, creating new collective works, for resale or redistribution to servers or lists, or reuse of any copyrighted component of this work in other works.
\end{minipage}}

\toappear{2024 IEEE International Workshop on Machine Learning for Signal Processing, Sept.\ 22--25, 2024, London, UK}

\usepackage{cite}
\usepackage{balance}

\usepackage{xcolor}

\usepackage{bm}
\def\vz{{\bm{z}}}
\def\vx{{\bm{x}}}

\def\vepsilon{{\boldsymbol{\epsilon}}}
\def\bs{\boldsymbol}

\usepackage{booktabs}
\usepackage{float}

\usepackage{enumitem}

\usepackage{amsmath,amsfonts,bm}

\def\eqref#1{equation~\ref{#1}}

\def\1{\bm{1}}

\def\rvm{{\mathbf{m}}}

\def\rvx{{\mathbf{x}}}

\def\vm{{\bm{m}}}

\def\vx{{\bm{x}}}

\def\vz{{\bm{z}}}

\DeclareMathAlphabet{\mathsfit}{\encodingdefault}{\sfdefault}{m}{sl}
\SetMathAlphabet{\mathsfit}{bold}{\encodingdefault}{\sfdefault}{bx}{n}

\title{WaveTransfer: A Flexible End-to-end Multi-instrument Timbre Transfer with Diffusion}
\name{Teysir Baoueb, Xiaoyu Bie, Hicham Janati, Gaël Richard \thanks{This work was funded by the European Union (ERC, HI-Audio, 101052978). Views and opinions expressed are however those of the author(s) only and do not necessarily reflect those of the European Union or the European Research Council. Neither the European Union nor the granting authority can be held responsible for them.}}
\address{LTCI, T\'el\'ecom Paris, IP Paris, France}

\begin{document}
\ninept

\maketitle

\begin{abstract}
As diffusion-based deep generative models gain prevalence, researchers are actively investigating their potential applications across various domains, including music synthesis and style alteration. Within this work, we are interested in timbre transfer, a process that involves seamlessly altering the instrumental characteristics of musical pieces while preserving essential musical elements. This paper introduces WaveTransfer, an end-to-end diffusion model designed for timbre transfer. We specifically employ the bilateral denoising diffusion model (BDDM) for noise scheduling search. Our model is capable of conducting timbre transfer between audio mixtures as well as individual instruments. Notably, it exhibits versatility in that it accommodates multiple types of timbre transfer between unique instrument pairs in a single model, eliminating the need for separate model training for each pairing. Furthermore, unlike recent works limited to $16$ kHz, WaveTransfer can be trained at various sampling rates, including the industry-standard $44.1$ kHz, a feature of particular interest to the music community.
\end{abstract}
\begin{keywords}
Multi-instrumental timbre transfer, diffusion models, music transformation, generative AI
\end{keywords}
\section{Introduction}\label{sec:introduction}
In recent years, there has been a growing interest in the manipulation and transformation of audio signals, particularly in the realm of music \cite{Brunner2018SymbolicMG, choi2020encoding, guo2022musiac}. One intriguing area of exploration within this domain is timbre transfer, a process that involves altering the tonal characteristics of musical sounds while preserving their content including fundamental pitch and temporal structure. Timbre is often described as the unique quality or color of a sound or `that attribute of auditory sensation in terms of which a listener can judge that two steady-state complex tones having the same loudness and pitch; are dissimilar' \cite{wallmark2022themeaning}. It plays a crucial role in shaping our perception and emotional response to music.
Timbre transfer can be considered as a more focused and well-defined objective than musical style transfer which usually involves not only timbre transfer (e.g., change of instrumentation) but also rhythmic and other high-level musical knowledge transfer (as in \cite{groove2groove} for music style transfer for symbolic music).

A large variety of approaches has already been proposed for timbre transfer.
Several methods rely on the modeling capabilities of autoencoders or Generative Adversarial Networks (GAN) to obtain a disentangled latent space suitable for timbre transfer. For instance, popular architectures include WaveNets autoencoders \cite{mor2018autoencoderbased}, Variational AutoEncoders (VAE) \cite{bitton2018modulated,cifka-VQVAE,Luo-GaussianVAE-2019} or GANs \cite{huang2018timbretron}. In \cite{Michelashvili2020HierarchicalTA}, an interesting hierarchical approach is proposed using source-filtering networks, which reconstruct the transferred signal at increasing resolution.

More recently, the potential of diffusion models for high-quality audio synthesis has opened a new path for diffusion-based timbre transfer.
For instance, in \cite{wu2023transplayer}, Transplayer utilizes a two-phase approach where the initial timbre transformation operated at the Constant Q Transform (CQT) representation level using an autoencoder architecture is further converted to audio waveform employing an audio-synthesis diffusion-based model. 
In \cite{Popov2023OptimalTI}, optimal transport principles are jointly exploited with diffusion modeling and successfully applied to the many-to-many timbre transfer task.

Other recent models for timbre transfer of particular interest for this work include the Music-STAR \cite{alinoori2022musicstar} and DiffTransfer\cite{comanducci2023timbre} systems.
Music-STAR \cite{alinoori2022musicstar} is built upon the WaveNet autoencoder \cite{engel2017neural} with a universal encoder and individual decoders corresponding to each of the target domains. 
In Difftransfer \cite{comanducci2023timbre}, the timbre transfer is carried out by means of Denoising Diffusion Implicit models \cite{song2021denoising}. This model was shown to be particularly efficient for both single- and multi-instrument timbre transfer and at the state of the art for the task on the Starnet dataset \cite{starnet}, also considered in this work.

In this paper, we introduce \textit{WaveTransfer}, a novel end-to-end diffusion model designed for timbre transfer. If our model shares some common concepts with the DiffTransfer model of \cite{comanducci2023timbre}, it is capable
of conducting timbre transfer between audio mixtures as well as individual instruments in a single global model eliminating the need for separate model training for each specific timbre transfer task. Another important property of our model is that it directly generates the audio waveform without needing to rely on an external vocoder. Finally, our model can operate at any sampling rate extending all previous works that are limited to a rather low $16$ kHz.

The paper will be structured as follows: Section \ref{sec:background} presents background work on denoising diffusion probabilistic models and bilateral denoising diffusion models. Section \ref{sec:proposed_method} introduces our timbre transfer method. Section \ref{sec:experiments} describes our experimental procedures, and Section \ref{sec:results} discusses the results. Finally, Section \ref{sec:conclusion} concludes with insights, a summary of our contributions, and suggestions for future research. Audio files and code are provided on our demo page: \href{https://wavetransfer.github.io/}{https://wavetransfer.github.io/}.

\begin{figure*}[!ht]
    \centering
    \includegraphics[width=0.7\textwidth]{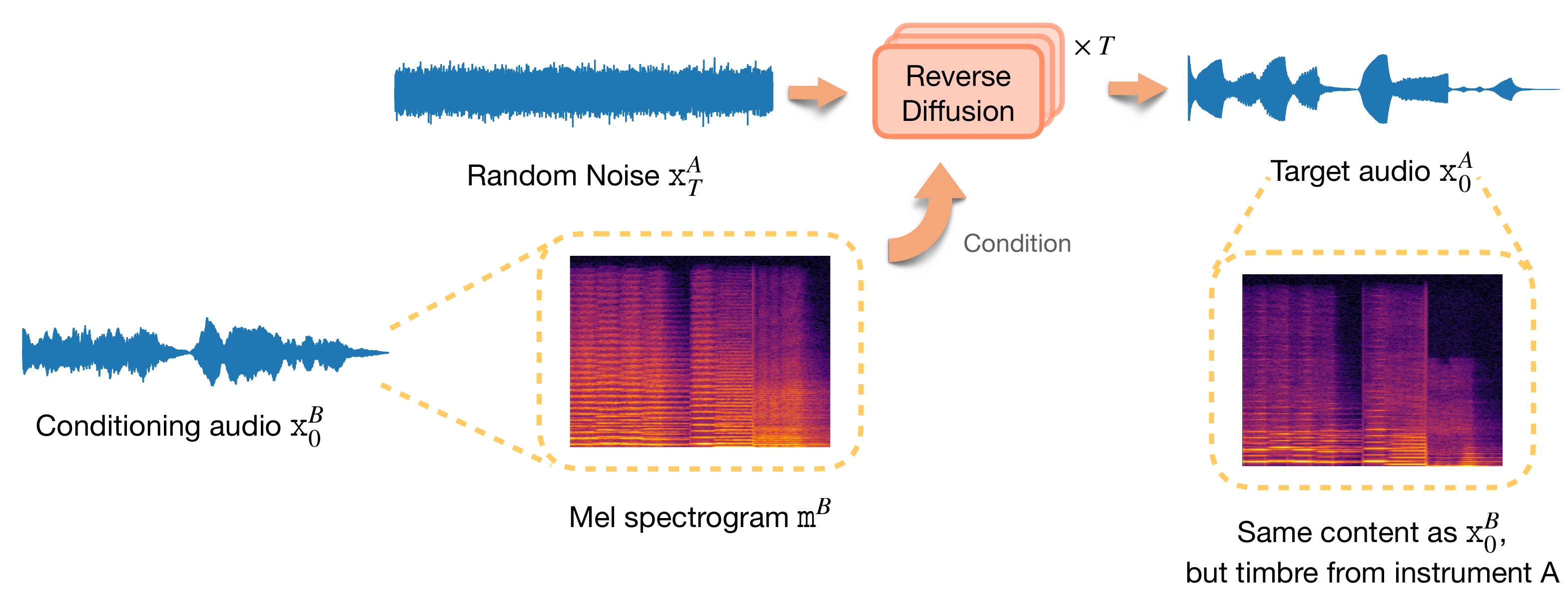}
    \caption{Timbre transfer using diffusion models. The objective is to generate a target audio $\rvx_0^A$ from a random noise $\rvx_T^A$ and a conditioning audio $\rvx_0^B$, where $\rvx_0^A$ has the same content as $\rvx_0^B$ but is played with a different instrument.}
    \label{fig:teaser}
\end{figure*}

\section{Background}
\label{sec:background}

\subsection{Denoising diffusion probabilistic models (DDPM)}
Denoising diffusion probabilistic models (DDPM) \cite{ho2020ddpm} represent a class of generative models characterized by a dual-process framework, the \textbf{forward} and \textbf{backward} processes.

In mathematical terms, consider $\rvx_0$ as a datum drawn from the distribution $q(\rvx_0)$ of a specified dataset. In the \textbf{forward} process, $\rvx_0$ is gradually perturbed by a Gaussian noise in $T$ steps, resulting in a sequence of noisy samples $\rvx_1, \dots, \rvx_T$. Given a noise schedule $\{\beta_t\}_{t=1}^T$, the forward diffusion process can be formulated as:
\begin{align}
q(\rvx_{t} | \rvx_{t-1}) &=\mathcal{N}(\rvx_{t}; \sqrt{1-\beta_{t}} \rvx_{t-1}, \beta_{t} \mathbf{I}).
\label{eqn:diff_proc_forward}
\end{align}
In a more concise manner, $\rvx_t$ can be sampled at any time step $t$ using the closed-form expression:
\begin{equation}
\rvx_t = \sqrt{\bar{\alpha}_t} \rvx_0 + \sqrt{1 - \bar{\alpha}_t} \vepsilon,
\label{eq:diff_proc_forward_compact}
\end{equation}
where $\vepsilon \sim \mathcal{N}(\mathbf{0}, \mathbf{I})$, $\alpha_t=1-\beta_t$ and $\bar{\alpha}_t = \prod_{i=1}^t \alpha_i$. When $T$ is sufficiently large,  $\rvx_T$ is equivalent to an isotropic Gaussian distribution.

If we can \textbf{reverse} the above process, we can generate new data from a Gaussian noise. However, directly computing the conditional distribution $q(\rvx_{t-1}|\rvx_{t})$ is not feasible, therefore we seek to learn a model  $p_{\bs{\theta}}(\rvx_{t-1}|\rvx_{t})$ that approximates the true distribution. The parameters $\bs{\theta}$ can be optimized by minimizing the Kullback-Leibler divergence between the two distributions, $\text{KL} (p_{\theta}(\rvx_{t-1}|\rvx_{t}) || q(\rvx_{t-1} | \rvx_{t}, \rvx_{0}))$. Since the reverse process is tractable conditioned on $\rvx_0$, we can obtain the analytical expression of $q(\rvx_{t-1} |\rvx_{t}, \rvx_{0})$ during training. In DDPM~\cite{ho2020ddpm}, the optimization is further simplified to the minimization of the noise estimation:
\begin{equation}
\mathcal{L}_{\bs{\theta}} = \min_{\bs{\theta}} \mathbb{E} \left[\left\|\vepsilon_\theta\left(\rvx_t, t\right)-\vepsilon\right\|_2^2\right],
\label{eq:diff_loss}
\end{equation}
where $t$ is the diffusion step randomly sampled from $[1, T]$, $\bs{\epsilon}$ is sampled from a normal distribution and $\rvx_t$ can be easily obtained via Eq. \ref{eq:diff_proc_forward_compact}. During inference, we can iteratively sample the data from $\rvx_T$ to $\rvx_0$ via\footnote{In DDPM \cite{ho2020ddpm}, both $\sigma_t^2=\beta_t$ and $\sigma_t^2 = \frac{1 - \bar{\alpha}_{t-1}}{1-\bar{\alpha}_{t}} \beta_t$ had similar results.}:
\begin{equation}
\rvx_{t-1} = \mathcal{N} \left(\rvx_{t-1}; \frac{1}{\sqrt{\alpha_t}} \left(\rvx_t - \frac{1-\alpha_t}{\sqrt{1-\bar{\alpha}}_t}\epsilon_{\theta}(\rvx_t, t)\right), \sigma_t^2 I\right)
\label{eq:diff_reverse}
\end{equation}

\subsection{Bilateral denoising diffusion models (BDDM)}
Given the premise of a sufficiently large value for $T$, executing the entire reverse process using Eq~\ref{eq:diff_reverse} is computationally expensive. To circumvent this, Lam et al. introduced bilateral denoising diffusion models (BDDM)~\cite{lam2022bddm}, an approach for judiciously determining an appropriate noise schedule with a length set to be within or match a specified maximum number of inference iterations.

More specifically, besides a diffusion model, an additional neural network, called the schedule network, is trained to select a noise schedule used for sampling at inference time. The training is done by minimizing the following objective function:
\begin{multline}
    \resizebox{\columnwidth}{!}{$\mathcal{L}^{(t)}_\phi =\frac{1}{2(1-\hat{\beta}_t(\phi) - \bar{\alpha}_t)} \left\lVert \sqrt{1-\bar{\alpha}_t} \vepsilon - \frac{\hat{\beta}_t(\phi)}{\sqrt{1-\bar{\alpha}_t}}\vepsilon_{\theta^\star}(\rvx_t, t) \right\rVert_2^2$}\\
    +\frac{1}{4} \log \frac{1-\bar{\alpha}_t}{\hat{\beta}_t(\phi)}+ \frac{D}{2}\left(\frac{\hat{\beta}_t(\phi)}{1-\bar{\alpha}_t}-1\right),
\end{multline}
where $\phi$ denotes the parameters of the schedule network and $\theta^\star$ represents well-optimized parameters of $\vepsilon_\theta$. Here $\hat{\beta}_t(\phi)$ denotes the noise scale at time $t$ which is obtained through the neural network of parameters $\phi$ by considering the previous noise scale $\hat{\beta}_{t+1}$ and the current noisy input $\rvx_t$.

\section{Proposed Method}\label{sec:proposed_method}
This section presents our approach for timbre transfer. Given a pair of tracks played with distinct instruments, our objective is to transfer the timbre from instrument $B$ (the conditioning instrument) to instrument $A$ (the target instrument), while maintaining the content from the conditioning instrument. As shown in Fig.~\ref{fig:teaser}, our model takes the mel spectrogram $\vm^B$ from the conditioning instrument $B$ as input, then applies an iterative diffusion process to generate a waveform $\vx_0^A$ with the timbre traits of the target instrument $A$. The objective of our model is to maximize the likelihood of the conditional distribution $q(\vx_0^A | \vm^B)$. 

\begin{figure*}[!ht]
    \centering
    \includegraphics[width=0.68\textwidth]{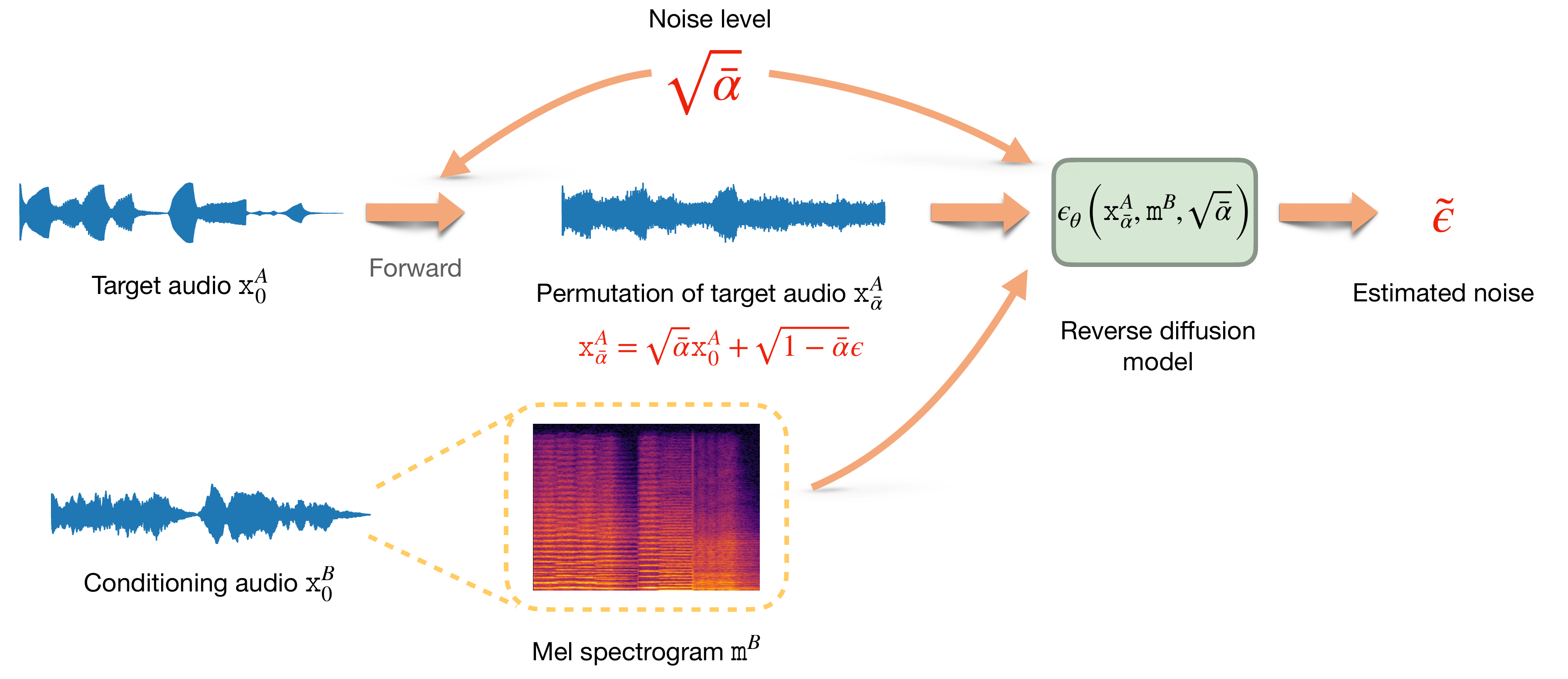}
    \caption{Training process of WaveTransfer}
    \label{fig:training}
\end{figure*}

\subsection{Training procedure}
The training process follows the principles of DDPM~\cite{ho2020ddpm}, and is depicted in Figure \ref{fig:training}. We start with an initial audio signal from the target instrument $\rvx_0^A \sim q_A(\rvx_0)$. Using the nice property from Eq. ~\ref{eq:diff_proc_forward_compact}, we can easily compute its perturbation $\rvx_t^A$ at diffusion step $t$. We then consider the corresponding audio $\rvx_0^B$ that has the same content as $\rvx_0^A$ but is played by a different instrument. Taking the mel spectrogram $\rvm^B$ of $\rvx_0^B$ as an additional condition, the model learns to predict the noise introduced to $\rvx_0^A$, leveraging the performance information encapsulated within $\rvm^B$.

Similar to WaveGrad~\cite{chen2021wavegrad}, we consider a continuous noise level $\sqrt{\bar{\alpha}}$ as conditioning provided to the neural network to serve the role of the time index $t$, where the sampling process for $\sqrt{\bar{\alpha}}$ involves utilizing a training noise schedule with length $T$. The training objective can thus be modified from Eq.~\ref{eq:diff_loss} as:
\begin{equation}
\mathcal{L}_{\bs{\theta}} = \min_{\bs{\theta}} \mathbb{E} \left[\left\|\vepsilon_\theta\left(\sqrt{\bar{\alpha}} \rvx_0^A + \sqrt{1-\bar{\alpha}}\vepsilon, \rvm^B,  \sqrt{\bar{\alpha}}\right)-\vepsilon\right\|_1\right],
\label{eq:timbre_loss}
\end{equation}

As previously mentioned, running inference with the $T$-long training noise schedule is computationally expensive. In WaveGrad~\cite{chen2021wavegrad}, Chen et al. proposed utilizing a grid search approach to select a shorter noise schedule. However, the search might take over a day for as few as $6$ iterations on $1$ NVIDIA Tesla P40 GPU, as observed by Lam et al. \cite{lam2022bddm}. Therefore, we opt to train a schedule network using the BDDM approach, as outlined in Section \ref{sec:background}, subsequent to training the timbre transfer neural network model.

\subsection{Model architecture}
The architecture of the timbre transfer neural network is similar to WaveGrad \cite{chen2021wavegrad}, featuring a series of upsampling blocks to expand the temporal dimension of the conditioning mel spectrogram $\rvm^B$ into the time domain. Conversely, downsampling blocks reduce the temporal dimension of the noisy audio input. Both pathways leverage a feature-wise linear modulation (FiLM) \cite{Perez2017FiLMVR} module to integrate the information gleaned from upsampling and downsampling processes synergistically.

The schedule network has a GALR (globally attentive locally recurrent) network architecture \cite{lam2021effective}. Within each GALR block, there exist two distinct modeling perspectives. The initial perspective focuses on recurrently modeling the local structures present in input signals, while the subsequent perspective is dedicated to capturing global dependencies through the utilization of the multi-head self-attention mechanism.

\subsection{Inference procedure}
\begin{figure}[!ht]
    \centering
    \includegraphics[width=0.98\columnwidth]{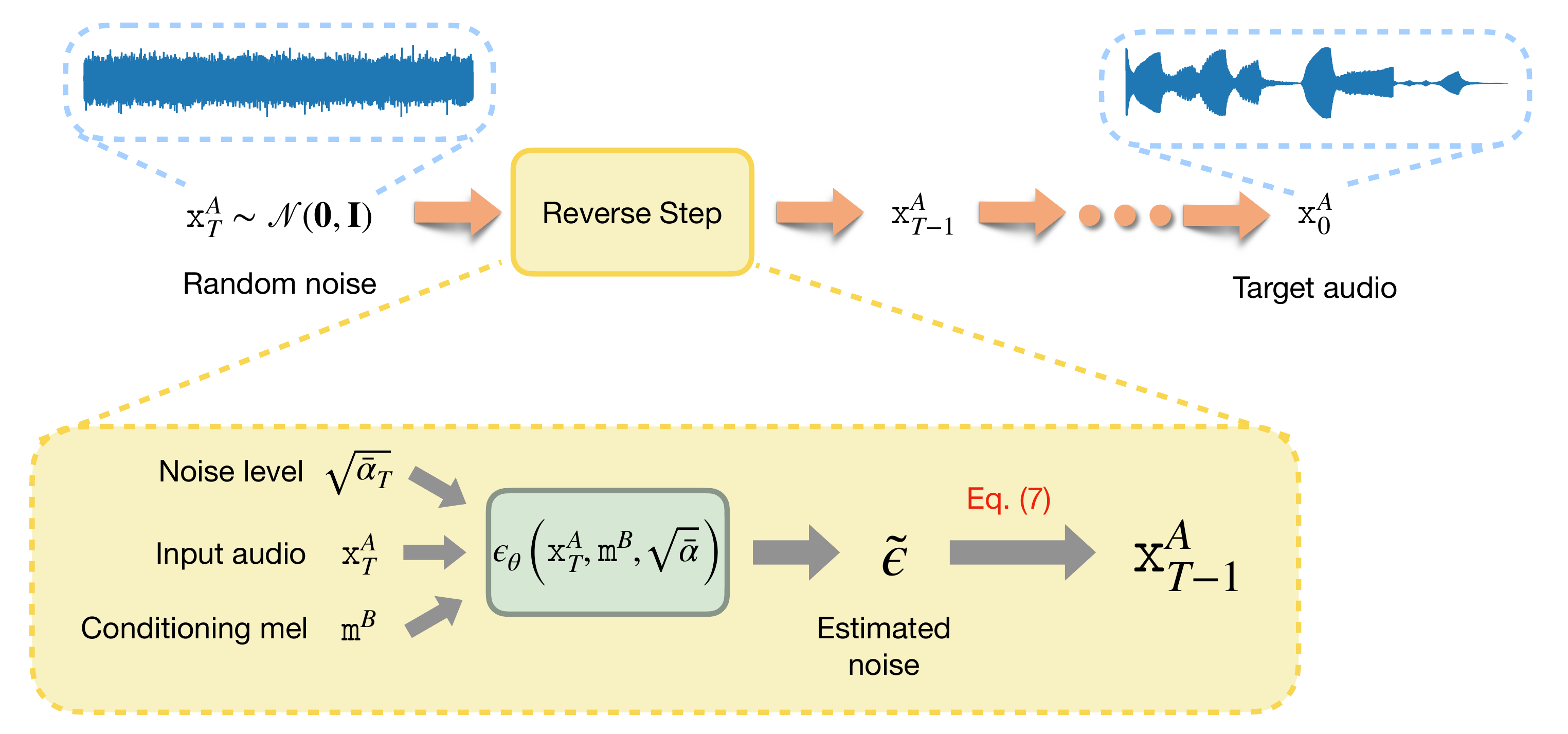}
    \caption{Inference process of WaveTransfer}
    \label{fig:inference}
\end{figure}

Given a noise schedule of length $N$, during inference, the model is provided with the mel spectrogram $\vm^B$ from the conditioning instrument $B$ along with random noise $\rvx_N \sim \mathcal{N}(\rvx_N; \mathbf{0}, \mathbf{I})$, as illustrated in Figure \ref{fig:inference}. The model approximates the added noise at each iteration. The estimated noise at step $n \in \![\![1, N\!]\!]$ is then used to generate $\rvx_{n-1}$. Finally, this iterative algorithm produces an audio signal with the same content as $\vm^B$ but played with instrument $A$. Similar to Eq.~\ref{eq:diff_reverse}, this procedure can be described by the following equation for DDPMs:
\begin{equation}
\rvx_{n-1}=\frac{1}{\sqrt{\alpha_n}}\left(\rvx_n-\frac{1-\alpha_n}{\sqrt{1-\bar{\alpha}_n}} \vepsilon_\theta\left(\rvx_n, \rvm^B, \sqrt{\bar{\alpha}_n}\right)\right) + \sigma_n \vz, \label{eq:inference}
\end{equation}
where $\vz \sim \mathcal{N}(\vz; \mathbf{0}, \mathbf{I})$ for $n > 1$, $\vz = \mathbf{0}$ for $n=1$ and $\sigma_n^2 = \frac{1 - \bar{\alpha}_{n - 1}}{1 - \bar{\alpha}_n}\beta_n$.

\section{Experiments}\label{sec:experiments}
In this section, we outline the experiments we conducted and the evaluation protocol employed.
\subsection{Dataset and preprocessing}\label{subsec:datasets}
The StarNet dataset \cite{starnet} consists of $104$ classical music compositions sampled at $44.1$ kHz. Each composition comprises six distinct tracks:
\begin{itemize}
\item $2$ Mixture tracks: clarinet-vibraphone, strings-piano
\item $4$ Individual stem tracks: clarinet, vibraphone, strings, piano
\end{itemize}
There is a correspondence between the content of the clarinet and strings tracks on one side and the piano and vibraphone tracks on the other.

The provided test set contains $10$ compositions, encompassing both classical and modern music pieces, with $6$ tracks each for stems and mixtures, resulting in a total of $60$ tracks.

In this work, we adopt the same preprocessing steps as in \cite{alinoori2022musicstar}, which involves detecting and removing intervals where one or both instruments are silent, and converting to mono for training on $44.1$ kHz tracks. For model training on $16$ kHz, we employ the reduced StarNet dataset, achieved through downsampling the preprocessed dataset. As the entire length of the track is not considered during training, but rather a random segment is extracted from each, we offline fragment the tracks into $5$-second excerpts for both training sets at each sampling rate to expedite loading time.

For validation purposes, we reserve $1$ composition ($\sim 100$ seconds $\times 6$) from the training set and utilize the remainder for training.

\subsection{Training setup}\label{subsec:training_configurations}
Four models were trained on the StarNet dataset, each serving different purposes.

The first two models perform all six timbre transfer types (clarinet $\leftrightarrow$ strings, piano $\leftrightarrow$ vibraphone, (clarinet + vibraphone) $\leftrightarrow$ (piano + strings)). These models, denoted as WT$_{\text{global}}^{16}$ and WT$_{\text{global}}^{44}$, were trained with sampling rates of $16$ kHz and $44.1$ kHz, respectively.

Furthermore, in order to evaluate the model's capability to execute timbre transfer between mixtures without the requirement of individual tracks, two additional models, WT$_{\text{mix}}^{16}$ ($16$ kHz) and WT$_{\text{mix}}^{44}$ ($44.1$ kHz), were trained exclusively on a subset of the training data containing only the two mixture tracks for each performance.

All models were trained on $1$ A$100$ GPU for $1$ M steps with a learning rate of $2 \cdot 10^{-4}$ and a batch size of $32$. We employ $128$-dimensional log-mel spectrograms calculated using a Hann window of $1200$ size, a hop length of $300$, and a $2048$-point FFT. We extract $66$ time frames from each training sample.

For each of the previous models, we train a schedule network for $10000$ steps on $1$ V$100$ GPU. We use $1$ GALR block with $128$ hidden dimensions, a window length of $8$ and a segment size of $64$.

\subsection{Metrics for evaluation}\label{subsec:metrics}
For objective evaluation, we employ the following metrics:
\begin{itemize}
    \item \textbf{Fréchet Audio Distance (FAD)} \cite{Kilgour2019FrchetAD} is a reference-free metric designed for evaluating audio quality. It utilizes a pre-trained audio model to generate embedding statistics for the set of produced tracks. These statistics are compared with those from a database of clean music by computing multivariate Gaussian distributions for each set of embeddings. The FAD score is then determined by calculating the Fréchet distance between these distributions. Smaller FAD scores indicate higher audio quality. We use the following models to compute the embeddings: the VGGish model \cite{vggish} ($16$ kHz, $44.1$ kHz), PANN \cite{Kong2019PANNsLP} ($16$ kHz) and CLAP \cite{Wu2022LargeScaleCL} ($44.1$ kHz)\footnote{Since VGGish and PANN are trained on $16$ kHz and CLAP is trained on  $48$ kHz, when we test waveforms at  $44.1$ kHz, we downsample them to $16$ kHz to compute VGGish embeddings and upsample to  $48$ kHz to compute the CLAP embeddings.}.
    \item \textbf{Perceptual Evaluation of Audio Quality (PEAQ)} \cite{peaq, vinay2023aquatk} is a standardized method used for evaluating the perceived audio quality of audio signals. It aims to quantify the difference between an original audio signal and a degraded version of that signal. It is composed of two scores:
    \begin{itemize}[topsep=0pt, itemsep=0pt, leftmargin=10pt]
        \item \textbf{Objective Difference Grade (ODG):} This metric quantifies the perceived quality difference between the original and generated signals. It assigns a score from $-4$ to $0$, where higher values indicate better quality.
        \item \textbf{Distortion Index (DI):} This index measures the level of distortion introduced in the generated signal. Lower values signify greater distortion.
    \end{itemize}
    \item \textbf{ViSQOL (Virtual Speech Quality Objective Listener)} \cite{chinen2020visqol} stands as a signal-based metric for full-reference assessment. Initially crafted to mirror human perception of speech quality, it relies on a spectro-temporal measure to gauge similarity between reference and test speech signals at 16 kHz. Subsequently, its scope expanded to encompass music signals at a 48 kHz sampling rate. When employing ViSQOL for evaluation, we upsample both the ground-truth and generated signals.
    
\end{itemize}

\subsection{Inference noise schedules}
During inference, we adopt WaveGrad's $6$-iteration noise schedule (WG-6). Additionally, given that DiffTransfer utilizes $20$ iterations for its reverse process, we investigate noise schedule searching with BDDM, setting a maximum of $20$ iterations. During the search, the network generates noise schedules of length $n \leq 20$. The BDDM approach necessitates employing a metric to determine the optimal noise schedule, evaluated based on its performance according to this metric on a validation set.

Given the slow computation of PEAQ and the potential inaccuracies in ViSQOL's assessment of short signals and its requirement of upsampling, we have decided to employ FAD alongside VGGish embeddings for this task, even though this choice is not flawless. One drawback lies in the necessity to compute the metric on a sufficiently large set of generated signals, which inevitably slows down the process compared to utilizing a rapid full-reference metric on as few as 1 sample as specified in \cite{lam2022bddm}. We denote the selected optimal noise schedule by BDDM-$n$.

\section{Results}\label{sec:results}
Hereafter, we present the results for timbre transfer, starting with the global models, which are capable of performing timbre transfer between individual stems and mixtures. Subsequently, we delve into the results concentrating on mixture timbre transfer, encompassing both global models and mixture-specific models.

\subsection{Inference conducted with global models}
In this subsection, we conduct the timbre transfer process across all $6$ possible transformations. To achieve this, we utilize both WT$_{\text{global}}^{16}$ and WT$_{\text{global}}^{44}$ on the $6$ tracks of the $10$ performances in the test set, resulting in a total of $60$ tracks. The results are presented in Tables \ref{tab:fad_1} and \ref{tab:full_reference_metrics_1}.

\begin{table}[h]
\caption{FAD results ($\downarrow$) on the test set ($60$ tracks) using $16$ kHz and $44.1$ kHz sampling rates and different embeddings}
\vspace{1pt}
 \label{tab:fad_1}
 \centering
 \resizebox{\columnwidth}{!}{
 \begin{tabular}{c|c|c|c|c}
 \toprule
  SR & Model & VGGish & PANN & CLAP\\ \midrule
  \multirow{2}{*}{\rotatebox[origin=c]{90}{\footnotesize{\textbf{16}}}} & WT$_{\text{global}}^{16}$ with BDDM-20 & $\mathbf{4.17}$ & $3.67 \cdot 10^{-3}$ & \textunderscore \\
  & WT$_{\text{global}}^{16}$ with WG-6 & $4.38$ & $\mathbf{3.59 \cdot 10^{-3}}$ & \textunderscore \\
  \midrule
  \multirow{2}{*}{\rotatebox[origin=c]{90}{\footnotesize{\textbf{44.1}}}} & WT$_{\text{global}}^{44}$ with BDDM-19 & $\mathbf{4.89}$ & \textunderscore & $\mathbf{0.51}$\\
  & WT$_{\text{global}}^{44}$ with WG-6 & $5.52$ & \textunderscore & $0.56$ \\
  \bottomrule
  \end{tabular}
  }
\end{table}

\begin{table}[h]
\caption{ViSQOL and PEAQ results (mean $\pm$ standard deviation) on the test set ($60$ tracks) using $16$ kHz and $44.1$ kHz sampling rates}
\vspace{1pt}
 \label{tab:full_reference_metrics_1}
 \centering
 \resizebox{\columnwidth}{!}{
 \begin{tabular}{c|c|c|c|c}
 \toprule
  SR & Model & ViSQOL ($\uparrow$) & ODG ($\uparrow$) & DI ($\uparrow$)\\ \midrule
  \multirow{2}{*}{\rotatebox[origin=c]{90}{{\textbf{16}}}} & WT$_{\text{global}}^{16}$ with BDDM-20 & $\mathbf{3.17} \pm 0.48$ & $-2.22 \pm 0.02$ & $-0.34 \pm 0.03$ \\
  & WT$_{\text{global}}^{16}$ with WG-6 & $3.13 \pm 0.53$ & $-2.22 \pm 0.02$ & $-0.34 \pm 0.03$ \\
  \midrule
  \multirow{2}{*}{\rotatebox[origin=c]{90}{{\textbf{44.1}}}} & WT$_{\text{global}}^{44}$ with BDDM-19 & $\mathbf{4.23} \pm 0.46$ & $-2.23 \pm 0.03$ & $-0.37 \pm 0.05$\\
  & WT$_{\text{global}}^{44}$ with WG-6 & $4.18 \pm 0.50$ & $-2.23 \pm 0.03$ & $-0.36 \pm 0.05$\\
  \bottomrule
  \end{tabular}
  }
\end{table}

We observe that employing the noise schedule derived from BDDM led to superior outcomes in terms of FAD scores with VGGish embeddings. This outcome aligns with expectations, as the selection of this particular noise schedule was predicated on its performance with respect to that metric. For the remaining embeddings, PANN shows comparable results, while CLAP exhibits a slight improvement with the BDDM approach.

Transitioning to full-reference metrics, we notice minimal variation in results between noise schedules, with the BDDM approach prevailing in ViSQOL but displaying almost no difference in PEAQ.

\subsection{Inference conducted only on mixture tracks}
In contradistinction to Models WT$_{\text{mix}}^{16}$ and WT$_{\text{mix}}^{44}$, DiffTransfer and Music-STAR train a single model for each type of transformation: one model for (piano + vibraphone) $\rightarrow$ (clarinet + vibraphone) and another one for (clarinet + vibraphone) $\rightarrow$ (piano + vibraphone). 

To ensure consistency with the evaluation protocols of DiffTransfer and Music-STAR in \cite{comanducci2023timbre}, we exclusively utilize the mixture tracks from each performance within the test set ($2$ per performance, totaling $20$ tracks). In addition to using WT$_{\text{mix}}^{16}$ and WT$_{\text{mix}}^{44}$, which are tailored explicitly for mixture-to-mixture timbre transfer, we incorporate models WT$_{\text{global}}^{16}$ and WT$_{\text{global}}^{44}$, where we focus only on evaluation with mixture tracks. The results~\footnote{The results reported for DiffTransfer and Music-STAR are extracted from \cite{comanducci2023timbre}. They were computed as follows: performing the timbre transfer task with each model: the (clarinet + vibraphone) $\rightarrow$ (piano + vibraphone) model and the (piano + vibraphone) $\rightarrow$ (clarinet + vibraphone) model, then running FAD on all generated mixture tracks.} are showcased in Tables \ref{tab:fad_2} 
and \ref{tab:full_reference_metrics_2}.

\begin{table}[h]
\caption{FAD results ($\downarrow$) on the mixture tracks in the test set ($20$ tracks) using $16$ kHz and $44.1$ kHz sampling rates and different embeddings}
\vspace{1pt}
 \label{tab:fad_2}
 \centering
 \resizebox{\columnwidth}{!}{
 \begin{tabular}{c|c|c|c|c}
 \toprule
  & Model & VGGish & PANN& CLAP\\ \midrule
  \multirow{6}{*}{\rotatebox[origin=c]{90}{\footnotesize{\textbf{16 kHz}}}} & DiffTransfer \cite{comanducci2023timbre} & $\mathbf{4.37}$ & $\mathbf{2.3 \cdot 10^{-3}}$& \textunderscore\\
  & Music-STAR \cite{alinoori2022musicstar} & $8.93$ & $3.3 \cdot 10^{-3}$ & \textunderscore\\
  & WT$_{\text{mix}}^{16}$ with WG-6 & $6.10$ & $3.90 \cdot 10^{-3}$ & \textunderscore\\
  & WT$_{\text{mix}}^{16}$ with BDDM-20 & $5.60$ & $3.42 \cdot 10^{-3}$ & \textunderscore\\
  & WT$_{\text{global}}^{16}$ with WG-6 & $6.34$ & $3.68 \cdot 10^{-3}$ & \textunderscore\\
  & WT$_{\text{global}}^{16}$ with BDDM-20 & $6.01$ & $3.75 \cdot 10^{-3}$ & \textunderscore\\
  \midrule
  \multirow{4}{*}{\rotatebox[origin=c]{90}{\footnotesize{\textbf{44.1 kHz}}}} & WT$_{\text{mix}}^{44}$ with WG-6 & $7.30$ & \textunderscore & $0.67$\\
  & WT$_{\text{mix}}^{44}$ with BDDM-20 & $6.74$ & \textunderscore & $\mathbf{0.63}$\\
  & WT$_{\text{global}}^{44}$ with WG-6 & $7.42$ & \textunderscore & $0.73$\\
  & WT$_{\text{global}}^{44}$ with BDDM-19 & $\mathbf{6.45}$ & \textunderscore & $0.67$\\
  \bottomrule
  \end{tabular}
  }
\end{table}

\begin{table}[h]
\caption{ViSQOL and PEAQ results on the mixture tracks in the test set ($20$ tracks) using $16$ kHz and $44.1$ kHz sampling rates}
\vspace{1pt}
 \label{tab:full_reference_metrics_2}
 \centering
 \resizebox{\columnwidth}{!}{
 \begin{tabular}{c|c|c|c|c}
 \toprule
  & Model & ViSQOL ($\uparrow$) & ODG ($\uparrow$) & DI ($\uparrow$)\\ \midrule
  \multirow{6}{*}{\rotatebox[origin=c]{90}{\footnotesize{\textbf{16 kHz}}}} & DiffTransfer \cite{comanducci2023timbre} & $\textbf{3.28} \pm 0.42$ & $\mathbf{-2.20} \pm 0.05$ & $\mathbf{-0.32} \pm 0.07$\\
  & Music-STAR \cite{alinoori2022musicstar} & $2.43 \pm 0.29$ & $-2.24 \pm 0.07$ & $-0.37 \pm 0.11$\\
  & WT$_{\text{mix}}^{16}$ with WG-6 & $3.02 \pm 0.32$ & $-2.22 \pm 0.02$ & $-0.34 \pm 0.03$\\
  & WT$_{\text{mix}}^{16}$ with BDDM-20 & $3.11 \pm 0.31$ & $-2.23 \pm 0.03$ & $-0.35 \pm 0.04$ \\
  & WT$_{\text{global}}^{16}$ with WG-6 & $2.86 \pm 0.33$ & $-2.22 \pm 0.03$ & $-0.34 \pm 0.03$ \\
  & WT$_{\text{global}}^{16}$ with BDDM-20 & $2.99 \pm 0.30$ & $-2.22 \pm 0.03$ & $-0.34 \pm 0.03$\\
  \midrule
  \multirow{4}{*}{\rotatebox[origin=c]{90}{\footnotesize{\textbf{44.1 kHz}}}} & WT$_{\text{mix}}^{44}$ with WG-6 & $3.98 \pm 0.58$ & $-2.24 \pm 0.03$ & $-0.37 \pm 0.05$ \\
  & WT$_{\text{mix}}^{44}$ with BDDM-20 & $2.82 \pm 0.83$ & $-2.25 \pm 0.04$ & $-0.40 \pm 0.08$\\
  & WT$_{\text{global}}^{44}$ with WG-6 & $3.76 \pm 0.71$ & $-2.24 \pm 0.03$ & $-0.37 \pm 0.05$\\
  & WT$_{\text{global}}^{44}$ with BDDM-19 & $\mathbf{4.06} \pm 0.54$ & $-2.24 \pm 0.04$ & $-0.38 \pm 0.06$\\
  \bottomrule
  \end{tabular}
  }
\end{table}

Once more, the noise schedule selected by BDDM demonstrates superior performance in FAD metrics when paired with VGGish and CLAP embeddings. However, the findings regarding PANN embeddings remain inconclusive, as the use of the BDDM-generated noise schedule sometimes leads to either improvement or deterioration. Additionally, enhancements in quality are evident with ViSQOL, yet outcomes with PEAQ lack decisiveness.

Comparing WT$_{\text{global}}^{16}$ with WT$_{\text{mix}}^{16}$ on one end, and WT$_{\text{global}}^{44}$ alongside WT$_{\text{mix}}^{44}$ on the other, showcases the efficacy of the approach without stems, eliminating the need for individual tracks featuring single instruments during timbre transfer within mixture compositions.

Compared to the baseline models, WaveTransfer surpasses Music-STAR across all metrics except FAD when utilizing PANN embeddings. Notably, WaveTransfer demonstrates smaller standard deviations, indicating a more consistent and stable generation process. Additionally, WaveTransfer performances approach those of DiffTransfer while our models employ a single model trained for both timbre transformations, contrasting with the need for separate models in DiffTransfer.

A subjective evaluation was performed using a MUSHRA test \cite{mushra}. The results, available on our demo page, clearly demonstrate the superiority of our model over the baseline models. The discrepancy with the objective FAD scores can be attributed to the fact that FAD scores do not consistently align with human perception. This misalignment has been noted in previous studies \cite{FAD1, FAD2}, where the choice of embedding significantly influences the results.

\subsection{Model complexity}
The WaveTransfer model has $15.92$ M parameters.

Concerning the time complexity, the WaveTransfer models trained at $16$ kHz generate at speeds of $\times 36.21$ times faster than real-time with a $6$-iteration noise schedule and $\times 9.65$ with a $20$-iteration schedule. Comparatively, WaveTransfer models trained at $44.1$ kHz demonstrate speeds of $\times 14.05$ ($6$-iteration noise schedule), $\times 3.93$ ($19$-iteration noise schedule), and $\times 3.72$ ($20$-iteration noise schedule) times faster than real-time.

\section{Conclusion \& Future Work}\label{sec:conclusion}
In this work, we introduced WaveTransfer, an end-to-end diffusion-based model designed for timbre transfer across both monophonic and polyphonic music, leveraging multi-instrument training. We effectively demonstrated the versatility and efficacy of our model by showcasing its performance across various sampling rates. Additionally, we incorporated the BDDM approach to enhance noise selection efficiency. By carefully choosing a fitting metric for noise schedule selection with BDDM, or by delving into alternative methods for determining inference noise schedules, we believe that there is ample room for enhancing the outcomes even further.

A constraint in our current methodology is the necessity for transferred timbre pairs to be disjoint. For example, if piano $\leftrightarrow$ vibraphone is a designated pair in the dataset, no other pair should involve either the piano or vibraphone. To address this limitation, our forthcoming efforts will focus on broadening the model's capabilities to encompass a wider array of instruments. This involves conditioning the network on instrument embeddings, enabling it to facilitate any-to-any timbre transfer.

\vfill
\pagebreak

\bibliographystyle{IEEEbib}
\bibliography{refs}

\end{document}